\documentclass[twocolumn,pre,floatfix]{revtex4}
\usepackage{psfig}
\usepackage{bm}

\newcommand{\kT}{k_{\rm B}T}
\newcommand{\dsig}{\delta\sigma}
\newcommand{\dlh}{\delta h}

\begin{document}
\title{Attractive instability of oppositely charged membranes
       induced by charge density fluctuations}
\author{Guy Hed} \author{S. A. Safran}
\affiliation{Department of Materials and Interfaces,
Weizmann Institute of Science, Rehovot 76100, Israel}

\begin{abstract}
We predict the conditions under which two oppositely charged
membranes show a dynamic, attractive instability. Two layers with
unequal charges of opposite sign can repel or be stable when in
close proximity. However, dynamic charge density
fluctuations can induce an attractive instability and thus facilitate fusion. We
predict the dominant instability modes and timescales and show
how these are controlled by the relative charge and membrane
viscosities. These  dynamic instabilities may be the precursors of
membrane fusion in systems where artificial vesicles are engulfed
by biological cells of opposite charge.
\end{abstract}
\maketitle

The interactions between lipid bilayers control many biological
processes such as membrane fusion, in which the domains separated
by each bilayer merge \cite{Sackmann95}. In order to fuse, the
bilayer must overcome an energetic barrier which can be reduced by
an increase of the local bilayer curvature \cite{Kozlovsky02}.
Thus, small wave length undulations of bilayers at close proximity
promote fusion. We show here that electrostatic interactions
between oppositely charged bilayers give rise to such fluctuations
and predict the dynamics of instabilities related to the coupled
distance and charge fluctuations.  These dynamics may control the
fusion of vesicles with oppositely charged cell membranes and our
results indicate how the time scale for such fusion can be
optimized as a function of the relative charges and viscosities.

The interaction of similarly charged surfaces is repulsive at the
mean-field level and in the limit of relatively high salt concentration
can be described  by the Debye-H\"uckel (DH) approximation to the
Poisson-Boltzmann equation \cite{Andelman95}. This repulsion
between the surfaces inhibits surface undulations and thus opposes
fusion, but fluctuations in the density of the charged lipids can sometimes
lead to an instability, as was demonstrated theoretically
\cite{Gelbart97} and numerically \cite{Kim03}.
The case of oppositely charged bilayers, however, is qualitatively different
because (as we show below) a system governed only by electrostatics
is {\em always} unstable; the unstable mode may be related to
charge fluctuations, height undulations, or a mixture of both,
depending on the spacing between the layers, the charge densities and the viscosities.

Recent experiments show that positively charged lipid-DNA
complexes fuse with negatively charged cell membranes \cite{Koltover98}.
The subsequent release and transfection of DNA in such systems
makes these complexes possible candidates for gene therapy.
Understanding and control of the fusion process in this system of
oppositely charged membranes is important so that one can minimize
the time in which the complex is engulfed by the cell.  In
addition, once the complex is inside the cell it is surrounded by
a positively charged bilayer that originated in the cell membrane.
Another fusion of these oppositely charged membranes is needed in
order to release the DNA into the cytoplasm \cite{Lin03}.
Our predictions for the stability of the system as a function of the
membrane charges and viscosities, may allow the optimization of
the charge and viscosity of the lipid-DNA complex in order to
enhance the fusion and transfection process.

Parsegian and Gingell \cite{Parsegian72} solved the
Poisson-Boltzmann equation using the DH approximation for two
oppositely charged surfaces. We  imagine that the charge density
is controlled by mixing charged and neutral lipids; this is the
case in both cellular membranes and in the lipid-DNA complexes.
When the charge densities on the surfaces are different, there are
counterions that must remain in the region between the surfaces in order to
balance the electrostatic interaction. At close proximity, the
pressure of these counterions leads to a net repulsion between the
surfaces. For flexible
surfaces, this short range repulsion inhibits undulations when the surfaces
are at a small distance. Such undulations are needed in order for
fusion to occur.  Thus, within this simple picture in which the
surface charges are uniform, oppositely charged surfaces do not
necessarily attract and will not necessarily fuse. In this paper,
we consider the effects of lateral fluctuations of the charges in
each membrane and show that at relatively short distances, these
fluctuations give rise to a dynamic instability that promotes
fusion. The instability involves a coupled mode of the local
charge densities and the distance between the bilayers.

In the DH limit, applicable to physiological salt concentrations,
the screening length that governs the interaction between the bilayers is $\kappa^{-1}$,
where $\kappa^2=\frac{8\pi n_{\rm s} Q_{\rm s}^2}{\epsilon \kT}$,
$n_{\rm{s}}$ is the salt concentration, $Q_{\rm{s}}$
is the charge per lipid and $\epsilon$ is the dielectric constant of water.
The charge density of each membrane $i=1,2$ is $\sigma_i e/a$,
where $a$ is the area of a lipid and $\sigma_i$ is a dimensionless
charge density. The electrostatic free energy of the system per
unit area is \cite{Parsegian72,Andelman95}
\begin{eqnarray}\label{eqfel}
&&f_{\rm el} =  \frac{2\pi e^2}{\epsilon\kappa a^2} \times \nonumber\\
&&\ \ \ \ \left( \frac{({\sigma_1}^2+{\sigma_2}^2)
\exp(- h) + 2\sigma_1\sigma_2} {\sinh(h)} + {\sigma_1}^2+{\sigma_2}^2 \right) ,\ \ \ \
\end{eqnarray}
where $h$ is the dimensionless distance between the membranes in units of
$\kappa^{-1}$. The last two terms in Eq. \ref{eqfel} account for
the electrostatic repulsion between lipids in the same bilayer.

We assume that the average distance between the bilayers is
imposed by external constraints, such as lateral or transversal
pressure, binding of trans-membranal proteins, or the system
geometry, as in the case where the cell membrane encloses the
lipid/DNA complex \cite{Lin03}. Similarly, the overall charge on
each membrane is conserved. We therefore focus only on local
changes in the inter-membrane spacing and charge density.

\begin{figure}[t]
\centerline{\psfig{figure=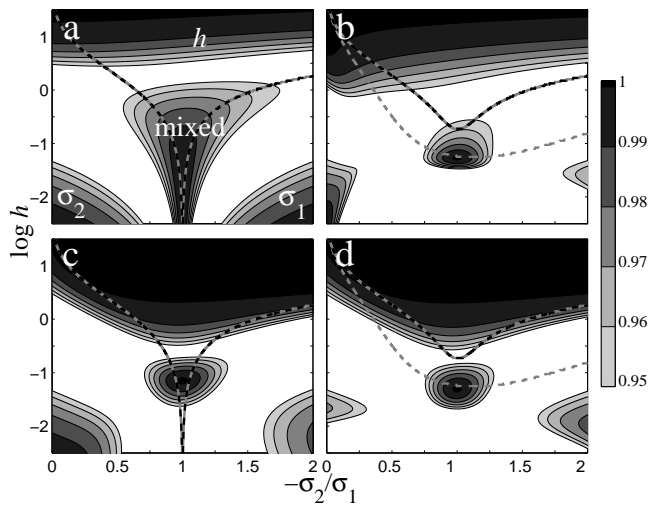,width=\columnwidth}} \caption{The
projections of the unstable eigenmodes of the system on the modes
predicted for the limiting cases. {\bf (a)} We use $f=f_{\rm el}$
and $\sigma_1=0.8$ to calculate the energy stability matrix, $\bm S$, at each point on the
$h\times\sigma_2$ surface. Let $\vec u$ be the eigenvector with
the negative eigenvalue of $\bm S$. In each region the color code
represents the projection $\vec u\cdot\vec v_i$, with $\vec v_i$
is $\vec v_h$, $\vec v_{\sigma_1}$, $\vec v_{\sigma_2}$ or $\vec
v_{\rm m}$ (defined in the text), according to the label on that
region. Below the black-grey dashed line $f_{hh}>0$ so a a system
with fixed constant charge density profile is stable there. {\bf
(b)} The same, but for $f=f_{\rm el}+f_{\rm ent}$. Below the dashed gray
line all the eigenvalues are positive, and there is no unstable
mode. We choose $\vec u$ there as the eigenvector with the
smallest eigenvalue. {\bf (c)} The same as in (a), but for the
unstable eigenvector of the dynamical matrix, ${\bm D}$, (see Eq. \ref{eqdyn}). {\bf (d)}
The same as in (b), but for the eigenvectors of ${\bm D}$.}
\label{fig1}
\end{figure}

As a first step, we consider only the electrostatic free energy,
that is, we let the total free energy, $f=f_{\rm el}$. In the calculations below, we
assume $\sigma_1>-\sigma_2>0$. The thermodynamic stability (see
below for the dynamics) of the system depends on the eigenvalues
of the Hessian matrix $\bm S$ of the second variations of the free
energy $f$,
\begin{equation}\label{eqS}
{\bm S} = \left( \begin{array}{ccc} f_{hh} & f_{h\sigma_1} & f_{h\sigma_2} \\
f_{h\sigma_1} & f_{\sigma_1\sigma_1} & f_{\sigma_1\sigma_2} \\
f_{h\sigma_2} & f_{\sigma_1\sigma_2} & f_{\sigma_2\sigma_2} \end{array} \right) \;.
\end{equation}
The determinant $|{\bm S}| = \left( \frac{4\pi e^2}{\epsilon\kappa
a^2} \right)^3 \frac{\sigma_1\sigma_2}{\sinh(h)}$ is always
negative for oppositely charged layers. In fact, $\bm S$ has one negative eigenvalue
$\lambda_i$, which corresponds to the unstable eigenmode $\vec v_i$
of the system.
We estimate this unstable mode analytically for three limiting cases:

{\bf (1)} For $h\gg 1$ we have $\lambda_h=\frac{8\pi
e^2}{\epsilon\kappa a^2} \sigma_1\sigma_2 \exp(-h)$ and $\vec
v_h=(1,0,0)$.  This represents a mode where there are surface
undulations  but with a constant charge density. This constant
density is the result of the repulsion between similar charges on
the same bilayer, which inhibits charge density fluctuations. In
this case, the instability is due to the ``naive'' picture of
attraction of oppositely (but uniformly) charged surfaces.

{\bf (2)} When $h\ll 1$ and $h\ll\sigma_1+\sigma_2$ the non-diagonal
terms of $\bm S$ vanish and $h$, $\sigma_1$ and $\sigma_2$
decouple. The unstable mode (under our assumption of $|\sigma_2|<|\sigma_1|$)
is $\vec v_{\sigma_2}=(0,0,1)$ with
$\lambda_{\sigma_2}=\frac{4\pi e^2}{\epsilon\kappa
a^2}\frac{\sigma_2}{\sigma_1+\sigma_2}h$ (in the case of
$|\sigma_2|>|\sigma_1|$ the unstable mode is $\vec
v_{\sigma_1}=(0,1,0)$). The associated density fluctuations tend
to lead to regions where locally $\sigma_1\approx -\sigma_2$
\cite{Nardi98}.  This is the situation in which the largest
attractions occur since there is no need for counterions to
balance the electrostatic interaction. Thus, the counterions may
be depleted from these regions and the associated entropic
repulsion between the layers vanishes. This situation, where the
interaction takes the form of a nearly bare Coulomb attraction
between surfaces, is energetically favorable. There are fluctuation
only in $\sigma_2$ since the repulsive interactions between
charges in the same bilayer inhibits charge fluctuations in the
more highly charged bilayer (${\sigma_2}^2<{\sigma_1}^2$) and the
charge fluctuations are therefore limited to the bilayer with the lower
charge density.

{\bf (3)} When $\sigma_1+\sigma_2 \ll h \ll 1$ the charge densities
are almost equal and opposite, $\sigma_1\approx -\sigma_2$, and
the interaction between the layers takes the form of an unscreened
Coulomb attraction. Charge density variations serve to increase
the instability. The eigenmode $\vec v_{\rm m} = \frac{1}{2}
\left( \sqrt{2},-1,1 \right)$ is an equal mixture of height
undulations and in-phase charge density fluctuations. The energy
associated with this instability is given by the eigenvalue
$\lambda_{\rm m} = -\frac{\sqrt{2}\pi e^2}{\epsilon\kappa
a^2}|\sigma_1-\sigma_2|$. In order to estimate the contribution of
the charge fluctuations to this instability, we compare
its energy scale, given by $\lambda_{\rm m}$, with
the energy scale for the same instability of a system with
uniform, local constant charge densities, given by
$f_{hh}$. To zeroth order in $\sigma_1+\sigma_2$ we have
$\frac{f_{hh}}{\lambda_{\rm m}} = \frac{\sigma_1\sinh(h/2)}{\sqrt{2}\cosh^3(h/2)}
\sim h \ll 1$.
This means that charge fluctuations, when they are allowed,
significantly increase the unstable eigenvalue and thus speed up the dynamical
attractive instability.

In order to more accurately determine the regions of validity of
these limiting cases, we calculated numerically the unstable
eigenmodes of the stability matrix, $\bm S$, and their projections
on each of the ``ideal'' modes found analytically in the limiting
cases discussed above. As seen in Fig. \ref{fig1}a, for $h\lesssim
1$ there is wide ranges of values of $\sigma_2$ where the mixed
mode $\vec v_{\rm m}$ is dominant. The charge density modes $\vec
v_{\sigma_1}$ and $\vec v_{\sigma_2}$ are dominant at $h\lesssim
0.25$ and $|1+\sigma_2/\sigma_1|>0.3$.

In addition to the electrostatic energy, charge fluctuations
caused by the demixing of the charged and neutral lipids also
modify the local entropy of each bilayer,
given by
$f_i = \frac{\kT}{a} \left[ c_i\log c_i + (1-c_i)\log(1-c_i)\right]$
(for $i=1,2$),
where $c_i=\sigma_i e/Q_i$ is the number density of the charged lipids
in bilayer $i$, and $Q_i$ is the charge of a single lipid.
The entropy will tend to
remix the two species and may inhibit to some degree or even
eliminate the charge instabilities that are promoted by the
electrostatic interactions. On the other hand, attractive interactions,
such as Van der Waals
attraction between similar lipids, promote the demixing charge
fluctuations and enhance the instability.
Here we consider the ``worst case'' where there are no
such attractions.

We also include the Helfrich repulsion between the bilayers, induced by
the entropy of the bilayer undulations \cite{Lipowsky95}.
Its contribution to the free energy is
$f_{\rm H} = c_{\rm H} (\kappa\kT)^2 (k_1^{-1}+k_2^{-1}) h^{-2}$, where
$c_{\rm H}\simeq 0.116$ is a universal number, and $k_i$ is the
bending modulus of bilayer $i$.
The contribution of the entropy of the charged lipids and of the undulations
to the free energy is $f_{\rm ent}=f_1+f_2+f_{\rm H}$.

In Fig. \ref{fig1}b, we present results for the eigenmodes of the
stability matrix, $\bm S$, for the total free energy $f=f_{\rm
el}+f_{\rm ent}$. The smallest eigenvalue is positive in the regions where
the $\vec v_{\sigma_1}$ and $\vec v_{\sigma_2}$ modes are
dominant, and the system is stable there. Nevertheless, there is a
finite range of $\sigma_2$ values, for which the mixed mode $\vec
v_{\rm m}$ is dominant, where the system is unstable even for
values of $h$ as small as 0.4.
Thus, the entropy modifies, but does not
eliminate entirely the predicted instabilities.

In order to test this prediction one can measure the stability of
multilayer systems made of
alternating layers of positive and negative charge. Systems that contain
random mixtures of
cationic and anionic lipids were already examined
\cite{Lewis00}. 
However, for the instabilities to be observed,
the cationic and anionic lipids must phase separate
(due to e.g., different chemistry of the polar heads or tails).
The spacing between the layers can be determined by the fraction of water in the
mixture. We predict that the system will be stable when the interlayer distance
$h$ is small, but it will become unstable when $h$ is larger than a critical
value $h_{\rm c}$ as shown in Fig. \ref{fig1},
which depends on the charge densities of the bilayers.

We now consider the dynamical response of the system  in
order to predict not just the regions of possible stability, but
also the time scales associated with the distance and charge
instabilities.  These timescales and their dependence on the
charge and the viscosities are important in controlling the fusion
processes in the experimental systems described above.  We
therefore consider the dynamical response of the system to  a
small perturbation with a long wavelength $2\pi q^{-1} \gg \bar h$, where
$\bar h=\kappa^{-1} h$ is the local instantaneous distance between the membranes,
and $q$ is the wave-number of the perturbation.

The hydrodynamics of the water layer between the bilayers in the
approximation of long wavelength perturbations is given, to first
order in $q \bar h$, by the lubrication approximation \cite{Oron97}:
\begin{equation}\label{eq3.1}
\eta\frac{\partial \bar h}{\partial t}= \frac{\partial}{\partial x} \left[
\frac{{\bar h}^3}{3} \frac{\partial}{\partial x} \left( \frac{\partial f}{\partial \bar h}
- \gamma \frac{\partial^2 \bar h}{\partial x^2} \right) \right] \;,
\end{equation}
where $\eta$ is the water viscosity and $\gamma$ is the membrane surface tension,
which we can neglect if the wave number $q$ is small enough
(see discussion at the end of the letter).

The dynamics of the charged lipids in the membrane is governed by
the Smoluchowski equation \cite{smolu}
\begin{equation}\label{eq3.2}
\frac{\partial c_i}{\partial t}=  \frac{\partial}{\partial x} \frac{1}{\zeta_i}
\left( c_i \frac{\partial \mu_i}{\partial x} \right) \,;\ \ \ \ \
\mu_i = \frac{\partial f}{\partial c_i} \;,
\end{equation}
where $\mu_i$ ($i=1,2$) is the chemical potential of a charged
lipids in bilayer $i$, $c_i$ is the number density of the charged
lipids and $\zeta_i$ is the two-dimensional viscosity in the
bilayer.

We consider a perturbation of the type
$h(x,t)=h_0+\delta h(t)\sin(\vec q \cdot \vec x)$,
$\sigma_i(x,t) = {\sigma_i}_0+\delta\sigma_i(t) \sin(\vec q \cdot \vec x)$.
From Eqs. \ref{eq3.1} and \ref{eq3.2} we obtain the dynamics to first order in
$\delta h$ and $\delta\sigma_i$:
\begin{eqnarray}\label{eqdyn}
\frac{\partial}{\partial t} \left(
\begin{array}{c} \dlh \\ \dsig_1 \\ \dsig_2 \end{array} \right) = -{\bm D}
\left( \begin{array}{c} \dlh \\ \dsig_1 \\ \dsig_2 \end{array} \right) ,
\;\;\;\;\;\;\;\;\;\;\;\;\;\;\; \nonumber\\
{\bm D} = q^2 \left(
\begin{array}{ccc} \frac{h^3}{3\eta\kappa}&0&0 \\ 0&\frac{Q_1 a\sigma_1}{e\zeta_1}&0\\
0&0&\frac{Q_2 a \sigma_2}{e \zeta_2} \end{array} \right) {\bm S} \;,
\end{eqnarray}
where the derivatives in $\bm S$ are evaluated at $h=h_0$ and ${\sigma_i}={\sigma_i}_0$.

\begin{figure}[t]
\centerline{\psfig{figure=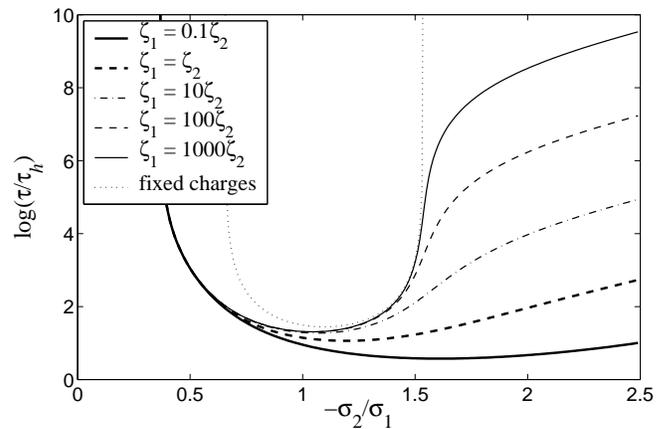,width=\columnwidth}}
\caption{The time scale $\tau$ for the growth of the instability for
different values of $\zeta_1$. $\tau$ is given by inverse of the negative
eigenvalue of the dynamical matrix, $\bm D$, calculated for
$f=f_{\rm el}+f_{\rm ent}$, $h=1$ and $\zeta_2=10^{-7}$ erg s/cm$^2$.
We keep $\sigma_1=0.8$ and vary $\sigma_2$ along the $x$-axis.
The dotted line is
the time scale for a system with a uniform charge density.
$\tau$ diverges when the system is stable.}
\label{fig2}
\end{figure}

From Eq. \ref{eqdyn} we obtain characteristic time scales for the
dynamics $h$, $\sigma_1$ and $\sigma_2$:
$\tau_h=  3\eta\kappa \frac{\epsilon\kappa a^2}{2\pi e^2} q^{-2}$,
$\tau_{\sigma_i}= \frac{\zeta_i}{Q_i}\frac{\epsilon\kappa a}{2\pi e} q^{-2}$.
For the numerical calculation of the time scales we use the
values: $\sigma_1=0.8$, $Q_1=-Q_2=2e$, $\eta=0.01$ erg s/cm$^3$,
$\zeta_1=\zeta_2=10^{-7}$ erg s/cm$^2$ \cite{Merkel89}, $T=298$ K,
$k_1\gg k_2=20\kT$,
$a=10^{-14}$ cm$^2$ and $\kappa=10^7$ cm$^{-1}$. With these values
we have $\tau_{\sigma_i}=35\tau_h$. If we consider
$\epsilon=4\pi\epsilon_0\epsilon_{\rm water}=3.5\cdot 10^{20}\ e^2$ erg$^{-1}$
cm$^{-1}$ and $q=10^5$ cm$^{-1}$ we have $\tau_h = 1.7\cdot
10^{-6}$ s.

The time scale for the growth of the instability is the inverse of
the negative eigenvalue of the dynamical matrix, ${\bm D}$ (in the
case that one exists). We have calculated the eigenmodes and time
scales for the two cases we considered: (i) electrostatics alone,
$f=f_{\rm el}$  (ii) electrostatics and entropy, $f=f_{\rm
el}+f_{\rm ent}$. In Figs. \ref{fig1}c and \ref{fig1}d we present the
regions where the unstable mode corresponds to one of the modes we
found above for the limiting cases. The figures demonstrate that
our results, obtained through analytical approximation, are
valid for large regions in the parameter space.

When we take into account the entropy (Fig. \ref{fig1}d) the mode
$\vec v_{\sigma_2}$ almost vanishes and the regions in which the modes
$\vec v_{\sigma_2}$ and $\vec v_{\sigma_1}$ were unstable in the absence of entropy,
become stable. However, there is
still a range of charge densities for which the system is
destabilized by charge fluctuations. This range is represented in
Fig. \ref{fig1}d by the area between the two stability lines. It
is mostly dominated by the mixed mode $\vec v_{\rm m}$ that
couples the charge density and distance fluctuations. This mixed mode results
in a time scale for the attractive instability that is significantly faster
than the time scale predicted from the mean-field attraction (i.e., in
the absence of charge density fluctuations).


In Fig. \ref{fig2} we present the dependence of the time scale $\tau$
for the instability on the intra-membrane viscosity $\zeta_1$.
As expected, we see that if $\zeta_1>\zeta_2$ the time
scales in the region where $|\sigma_2|>|\sigma_1|$, in which $\vec
v_{\sigma_1}$ mode is dominant, are much slower than the time
scales in the opposite region, where $\vec v_{\sigma_2}$ mode is
dominant, and vice versa.
An increase in the viscosity $\zeta_1$ shifts the
maximal instability (and shortest time scale) to lower values of the charge ratio.
This prediction indicates how one might optimize of the charge and viscosity values
in the fusion and transfection experiments.

The electrostatic free energy $f_{\rm el}$ promotes that growth of
perturbations with any wave number $q$. These instabilities in
 are inhibited by the surface tension $\gamma_i$
and the bending modulus $k_i$ of each membrane labelled by $i=1,2$.
For simplicity, we assume here that
$k_2=k\ll k_1$ and $\gamma_2=\gamma\ll\gamma_1$, so that
variations of the inter-membrane distance, $h$, only arise from
undulations of the more flexible bilayer, denoted as bilayer 2.
In this case, the only contribution of these
terms to the stability matrix, $\bm S$, (defined for the
electrostatic-only case, $f=f_{\rm el}$) will be in the entry
$S_{11} = {f_{\rm el}}_{hh} + \gamma q^2 + \frac{k}{2} q^4$, and
$|{\bm S}| = \frac{16\pi^2 e^4}{\epsilon^2\kappa^2 a^4}
\left(\frac{4\pi\sigma_1\sigma_2 e^2}{\epsilon\kappa a^2 \sinh(h)}
+ \gamma q^2 + \frac{k}{2} q^4\right)$.

Let $q_{\rm max}$ be the wave number at which the system is
marginally stable:  $|{\bm S}|=0$. The system is unstable for all
perturbations with longer wavelengths, $q<q_{\rm max}$. From Eq.
\ref{eqdyn} one sees that the time scale for the increase of a
perturbation is proportional to $q^{-2}$. Thus the dominant
instability of the system will be with a wave number $q\simeq
q_{\rm max}$.

At a small inter-membrane separation $h\ll 1$ we have $q_{\rm max}\sim h^{-\frac{1}{4}}$.
As $h\to 0$ our theory predicts the growth of unstable modes with large wave vectors.
The related undulations will grow until the bilayers are at a microscopic distance,
where the electrostatic instability is balanced by short-range hydration repulsion.
Thus, there will be regions where the bilayers are at close proximity and have
high local curvature. In these regions the barrier for fusion is reduced considerably
\cite{Kozlovsky02}, and the probability for fusion is greatly increased.

The authors acknowledge useful discussions with C. R. Safinya, V. A. Parsegian
and N. Kampf as well as the support of the Israel Science Foundation.


\newpage

\end{document}